\documentstyle[12pt,epsfig]{article}

\newskip\humongous \humongous=0pt plus 1000pt minus 1000pt

\newif\ifdtup

\def\pr#1{#1^\prime}

\def\beq{\begin{equation}}
\def\eeq{\end{equation}}

\def\beqn{\begin{eqnarray}}
\def\eeqn{\end{eqnarray}}
\relax

\def\dotx{\dotx{\dot\overline{x}}}

\relax

\catcode`\@=11

\@addtoreset{equation}{section}
\def\theequation{\thesection\arabic{equation}}

\def\@normalsize{\@setsize\normalsize{15pt}\xiipt\@xiipt
\abovedisplayskip 14pt plus3pt minus3pt%
\belowdisplayskip \abovedisplayskip
\abovedisplayshortskip \z@ plus3pt%
\belowdisplayshortskip 7pt plus3.5pt minus0pt}

\def\small{\@setsize\small{13.6pt}\xipt\@xipt
\abovedisplayskip 13pt plus3pt minus3pt%
\belowdisplayskip \abovedisplayskip
\abovedisplayshortskip \z@ plus3pt%
\belowdisplayshortskip 7pt plus3.5pt minus0pt
\def\@listi{\parsep 4.5pt plus 2pt minus 1pt
     \itemsep \parsep
     \topsep 9pt plus 3pt minus 3pt}}

\@twosidetrue

\relax

\catcode`@=12

\catcode`\@=11

\def\section{\@startsection{section}{1}{\z@}{3.5ex plus 1ex minus
   .2ex}{2.3ex plus .2ex}{\large\bf}}

\def\thesection{\arabic{section}.}

\def\appendix{\setcounter{section}{0}
 \def\thesection{APPENDIX \Alph{section}:}
 \def\theequation{\Alph{section}.\arabic{equation}}}

\def\ps@headings{\def\@oddfoot{}\def\@evenfoot{}
\def\@oddhead{\hbox{}\hfill
 \makebox[.5\textwidth]{\raggedright\ignorespaces --\thepage{}--
 \hfill {}}}  
\def\@evenhead{\@oddhead}
\def\subsectionmark##1{\markboth{##1}{}}
}

\ps@headings

\catcode`\@=12

\def\figcap{\section*{Figure Captions\markboth
 {FIGURECAPTIONS}{FIGURECAPTIONS}}\list
 {Fig. \arabic{enumi}:\hfill}{\settowidth\labelwidth{Fig. 999:}
 \leftmargin\labelwidth
 \advance\leftmargin\labelsep\usecounter{enumi}}}
 \relax
\def\tablecap{\section*{Table Captions\markboth
 {TABLECAPTIONS}{TABLECAPTIONS}}\list
 {Table \arabic{enumi}:\hfill}{\settowidth\labelwidth{Table 999:}
 \leftmargin\labelwidth
 \advance\leftmargin\labelsep\usecounter{enumi}}}
 \relax
\def\reflist{\section*{References\markboth
 {REFLIST}{REFLIST}}\list
 {[\arabic{enumi}]\hfill}{\settowidth\labelwidth{[999]}
 \leftmargin\labelwidth
 \advance\leftmargin\labelsep\usecounter{enumi}}}
 \relax

\catcode`\@=11

\def\ps@headings{\def\@oddfoot{}\def\@evenfoot{}
\def\@oddhead{\hbox{}\hfill
 \makebox[.5\textwidth]{\raggedright\ignorespaces --\thepage{}--
 \hfill {}}}    
\def\@evenhead{\@oddhead}
\def\subsectionmark##1{\markboth{##1}{}}
}

\ps@headings

\catcode`\@=12

\relax

\catcode`\@=11
\def\prm{\fam \z@}
\catcode`\@=12

\relax
\def\pl#1#2#3{{\it Phys. Lett. }{\bf #1}(19#2)#3}
\def\zp#1#2#3{{\it Z. Phys. }{\bf #1}(19#2)#3}

\def\pr#1#2#3{{\it Phys. Rev. }{\bf #1}(19#2)#3}
\def\np#1#2#3{{\it Nucl. Phys. }{\bf #1}(19#2)#3}

\def    \hepph  #1 {{\tt hep-ph/#1}}
\def    \hepex  #1 {{\tt hep-ex/#1}}

\relax

\begin{document}            
\newcommand\SS{\scriptsize}
\newcommand\sss{\scriptscriptstyle}
\newcommand\gs{g_{\sss S}}
\newcommand\as{\alpha_{\sss S}}         
\newcommand\ep{\epsilon}
\newcommand\Th{\theta}
\newcommand\epb{\overline{\epsilon}}
\newcommand\kph{k_\gamma}
\newcommand\ki{k_i}
\newcommand\qa{q_a}
\newcommand\epem{e^+e^-}
\newcommand\Riph{R_{i\gamma}}
\newcommand\Rjph{R_{j\gamma}}
\newcommand\Rroph{R_{\rho\gamma}}
\newcommand\Xfun{{\cal X}}
\newcommand\dzero{\delta_0}
\newcommand\epph{\epsilon_\gamma}
\newcommand\th{\theta}

\newcommand\MSB{{\rm \overline{MS}}}
\newcommand\DIG{{\rm DIS}_\gamma}
\newcommand\CA{C_{\sss A}}
\newcommand\DA{D_{\sss A}}
\newcommand\CF{C_{\sss F}}
\newcommand\TF{T_{\sss F}}
\newcommand\SVfact{\frac{(4\pi)^\ep}{\Gamma(1-\ep)}
                   \left(\frac{\mu^2}{Q^2}\right)^\ep}
\renewcommand\topfraction{1}       
\renewcommand\bottomfraction{1}    
\renewcommand\textfraction{0}      
\setcounter{topnumber}{5}          
\setcounter{bottomnumber}{5}       
\setcounter{totalnumber}{5}        
\setcounter{dbltopnumber}{2}       
\newsavebox\tmpfig
\newcommand\settmpfig[1]{\sbox{\tmpfig}{\mbox{\ref{#1}}}}
%
\begin{titlepage}
\nopagebreak
{\flushright{
        \begin{minipage}{4cm}
        ETH-TH/97-40 \hfill \\
        hep-ph/yymmxxx\hfill \\
        \end{minipage}        }

}
\vfill
\begin{center}

{\large {\sc Isolated photons\\[0.3cm]
             in perturbative QCD}} 
\vskip .5cm
{\bf Stefano FRIXIONE}\footnote{Work supported by the Swiss National
Foundation.}
\\                    
\vskip .1cm
{Theoretical Physics, ETH, Zurich, Switzerland} \\
\end{center}
\nopagebreak
\vfill
\begin{abstract}
I present a definition of the cross section for the 
production of an isolated photon plus $n$ jets
which only depends upon direct photon production, and it is 
independent of the parton-to-photon fragmentation contribution.
This prescription, based on a modified cone approach which implements
the isolation condition in a smooth way, treats in the same way quarks
and gluons and can be directly applied to experimental data in hadron-hadron,
photon-hadron and $\epem$ collisions. The case of several, isolated
photons in the final state can also be dealt with in the very same way.

\end{abstract}        
\vfill
\end{titlepage}

\section{Introduction}

Photons are produced in scattering phenomena by two different
mechanisms. In the direct process, the photon enters the partonic 
hard collision, characterized by a large energy scale. In the
fragmentation process, a QCD parton (quark or gluon) fragments
non-perturbatively into a photon, at a scale of the order of the
typical hadronic mass. The former process is computable in
perturbative QCD, while the latter is not; all the unknowns of
the fragmentation mechanism are collected into two functions
(the quark-to-photon and gluon-to-photon fragmentation functions)
which, although universal, must be determined by comparison with
the data. Direct photons are usually well isolated from 
the final state hadrons, while photons produced via
fragmentation usually lie inside hadronic jets. From the experimental
point of view, it is not difficult to select a data sample in which 
the direct mechanism is dominant over the fragmentation mechanism,
by rejecting all those events where the photon is not isolated 
from hadronic tracks. A clean sample of well-isolated 
photons is extremely useful for a variety of topics,
like a detailed understanding of the underlying parton picture,
to constrain the gluon density of the proton in an intermediate $x$ range,
and to obtain an efficient background rejection in Higgs searches
at future colliders.

In perturbative QCD, the problem is more involved. It is not possible
to separate sharply the photon from the partons; in fact, this would 
constrain the phase space of soft gluons, thus spoiling the cancellation
of infrared divergences which is crucial in order to get a sensible
cross section. Two methods have been devised to tackle this problem. 
In the cone approach~[\ref{cone}], a cone 
is drawn around the photon axis; if only a small hadronic
energy (compared to the photon energy) is found inside the cone,
the partons accompanying the photon are clustered with a given jet-finding
algorithm. In the democratic approach~[\ref{demo}] the photon 
is treated as a parton as far as the jet-finding algorithm is
concerned. At the end of the clustering procedure, the configuration
corresponds to an isolated photon event only if the ratio of the
hadronic energy found inside the jet containing the photon over the total 
energy of the jet itself is smaller than a fixed amount, usually less 
than 10\%. The democratic approach is more suited than the cone approach
to extract the non-perturbative parton-to-photon fragmentation functions 
from the data. This point of view was adopted in ref.~[\ref{fragfun}].

In this paper, I will deal with the problem of defining an isolated-photon
cross section which minimizes the contribution of the fragmentation
mechanism. In particular, I will show that it is possible to modify
the cone approach in order to get a cross section which {\it only}
depends upon the direct process. I argue that this prescription
is infrared safe at any order in perturbative QCD. The paper is organized 
as follows: in section 2 I sketch the main ideas in a simple way. 
In section 3 I give a precise definition of the isolation conditions. 
Finally, section 4 contains my conclusions.

\section{A simple formulation of the problem}

In the cone approach, the first naive procedure is to draw a cone 
around the photon axis, and to impose that no quark or gluon is found 
inside the cone. With this definition, the configurations where a 
parton is collinear to the photon are rejected, and therefore the
contribution of the fragmentation process is exactly zero.
Unfortunately, this prescription is not infrared safe: soft
gluons can not be emitted inside the cone, thus spoiling the
cancellation of infrared singularities. One may relax the definition,
by allowing a small amount of hadronic energy inside the cone.
This restores the correct infrared behaviour, 
but at the same time introduces a dependence
upon the fragmentation functions, since collinear configurations
are not forbidden any longer. Another possibility, which only works
at next-to-leading order, is to allow soft gluons inside the 
cone, but to exclude the quarks. This prescription, however, 
is non-physical and its predictions can not be straightforwardly 
compared with experimental results. 

Therefore, to define an infrared-safe cross section, there should be no 
region of forbidden radiation in the phase space, while to eliminate the
dependence upon the fragmentation functions such a region must exist;
these two requirements are seemingly incompatible. I will argue in
the following that actually this is not the case. In fact, the
fragmentation mechanism in QCD is a purely collinear phenomenon;
therefore, to eliminate its contribution to the cross section, it
is sufficient to veto the collinear configurations only. In practice,
this can be achieved in the following way (I restrict for the moment 
to the case of $\epem$ collisions). A cone of (fixed) half-angle
$\dzero$ is drawn around the photon axis. Then, {\it for all}
\mbox{$\delta\le\dzero$}, the total amount of hadronic energy 
$E_{tot}(\delta)$ found inside the cone of half-angle $\delta$ drawn 
around the photon axis is required to fulfill the following condition
\beq
E_{tot}(\delta)\,\le\,{\cal K}\,\delta^2,
\label{iscond0}
\eeq
where ${\cal K}$ is some energy scale (the form ${\cal K}\,\delta^2$
is chosen for illustrative purposes; it will be generalized 
in the following). According to eq.~(\ref{iscond0}), a soft 
gluon can be arbitrarily close to the photon. On the other hand,
eq.~(\ref{iscond0}) implies that the energy of a parton emitted exactly 
collinear to the photon must vanish. Therefore, the contribution of the 
fragmentation process is restricted to the zero-measure set $z=1$.

In the following section, I will give a precise definition of the 
isolation condition, refining eq.~(\ref{iscond0}). Here, I stress that
eq.~(\ref{iscond0}) does not spoil the cancellation of soft
gluon effects, and achieves the isolation of the photon in a smooth
way, which can be easily implemented at the experimental level.

\section{Isolated-photon plus jets cross section}

I start from the class of scattering events whose final state contains 
a set of hadrons, labelled by the index $i$, with four-momenta $\ki$, 
and a hard photon with four-momentum $\kph$. I assume to be in a 
kinematic regime where the masses of the hadrons are small 
compared to their (transverse) energies. Also, in a real experimental 
situation, we may think of the $\ki$ as the four momenta deposited
in the $i^{th}$ calorimetric cell, instead of the four momentum
of the $i^{th}$ hadron. Fix the parameter $\dzero$, which defines
the so-called isolation cone, and apply to each event the following 
procedure ({\it isolation cuts}).

\begin{enumerate}

\item For each $i$, evaluate the angular distance $\Riph$ between
$i$ and the photon. The angular distance is defined, in the case of 
$\epem$ collisions, to be
\beq
\Riph=\delta_{i\gamma},
\label{Repem}
\eeq
where $\delta_{i\gamma}$ is the angle between the three-momenta of $i$ 
and $\gamma$. In the case of hadronic collisions I define instead
\beq
\Riph=\sqrt{(\eta_i-\eta_\gamma)^2+(\varphi_i-\varphi_\gamma)^2},
\label{Rhad}
\eeq
where $\eta$ and $\varphi$ are the pseudorapidity and azimuthal angle 
respectively.

\item Reject the event unless the following condition is fulfilled
\beq
\sum_i E_i\,\Th(\delta-\Riph)\,\le\,\Xfun(\delta)\;\;\;\;\;\;
{\rm for~all}\;\;\;\;\;\;\delta\le\dzero,
\label{iscond}
\eeq
where $E_i$ is the energy of hadron $i$ and, due to 
\mbox{$\Th(\delta-\Riph)$}, the sum gets contribution only from
those hadrons whose angular distance from the photon is smaller than 
or equal to $\delta$. The function $\Xfun$, which plays the
r\^{o}le of \mbox{${\cal K}\delta^2$} in eq.~(\ref{iscond0}), is 
fixed and will be given in the following. The function $\Xfun$ must
vanish when its argument tends to zero, \mbox{$\Xfun(\delta)\to 0$} 
for \mbox{$\delta\to 0$}. At hadron colliders, the transverse energy 
$E_{i{\sss T}}$ must be used instead of $E_i$.

\item Apply a jet-finding algorithm to the hadrons of the event
(therefore, the photon is excluded). This will result in a set of 
$m+m^\prime$ bunches of well-collimated hadrons, which I denote as 
candidate jets. $m$ ($m^\prime$) is the number of candidate jets which 
lie outside (inside) the isolation cone, in the sense of the angular 
distance defined by eqs.~(\ref{Repem}) or~(\ref{Rhad}). 

\item Apply any other additional cuts to the photon and to
the $m$ candidate jets which lie outside the cone (for example, the cut 
over the minimum observable (transverse) energy of the jets must be 
applied here). 

\end{enumerate}

An event which is not rejected when the isolation cuts are
applied is by definition an {\it isolated-photon plus $m$-jet} event.
The key point in the above procedure is step 2: hadrons are allowed inside 
the isolation cone, provided that eq.~(\ref{iscond}) is fulfilled. 
This in turn implies the possibility for a candidate jet to be
inside the isolation cone. It would not make much sense to define
a cross section exclusive in the variables of such a jet, which can
not be too hard. For this reason, in the physical observable that I 
define here, the jets which accompany the photon are the candidate 
jets outside the isolation cone which also pass the
cuts of step 4. The resulting cross section is therefore totally
exclusive in the variables of these jets and of the photon, and
inclusive in the variables of the hadrons found inside the isolation cone.
Notice that this is not equivalent to applying a jet-finding algorithm
only to the hadrons lying outside the isolation cone; in fact,
such a procedure is not infrared-safe.

I define
\beq
\Xfun(\delta)=E_\gamma\epph
\left(\frac{1-\cos\delta}{1-\cos\dzero}\right)^n,
\label{Xfundef}
\eeq
where $E_\gamma$ is the photon energy (in the case of hadron 
collisions, $E_\gamma$ must be replaced by the transverse energy of
the photon, $E_{\gamma {\sss T}}$). I will use
\beq
\epph=1,\;\;\;\;n=1.
\label{parameters}
\eeq
The reason for this choice will be discussed in the following. 
Here, I stress that this choice is arbitrary to a large extent. 
The main feature of the function $\Xfun$ is that
\beq
\lim_{\delta\to 0}\,\Xfun(\delta)=0.
\label{Xfunlim}
\eeq

In QCD, any jet cross section is easily written in terms of
measurement functions~[\ref{KS}]. Given a $N$-parton configuration
$\{\ki\}_{i=1}^N$, the application of a jet-finding algorithm results in
a set of $M$ jets with momenta $\{\qa\}_{a=1}^M$. This can be formally
expressed by the measurement function
\beq
{\cal S}_N\left(\{q_a\}_{a=1}^M;\{\ki\}_{i=1}^N\right),
\label{measfun}
\eeq
which embeds the definition of the jet four-momenta in 
terms of the parton four-momenta. It has been 
shown~[\ref{KS},\ref{FKS},\ref{Jets97}] that, at next-to-leading 
order and for an arbitrary type of collisions, the infrared-safeness 
requirement on the cross section can be formulated in terms of conditions
relating the measurement functions ${\cal S}_N$ for different $N$
(I refer the reader to the original publications for details).
These conditions can be extended without any difficulties to higher
perturbative orders. Here, I stress that the measurement function in 
eq.~(\ref{measfun}) implements an infrared-safe jet cross section 
definition, which I will apply to the partons accompanying the photon 
in a candidate isolated-photon event. By labeling the partons
in such a way that
\beq
\Riph\,\ge\,\Rjph\;\;\;\;\;\;
{\rm if}\;\;\;\;\;\; i\,>\,j,
\eeq
I define
\beqn
&&{\cal S}_{\gamma,N}\left(\kph,\{q_a\}_{a=1}^M;\{\ki\}_{i=1}^N\right)=
{\cal S}_N\left(\{q_a\}_{a=1}^M;\{\ki\}_{i=1}^N\right)
\times \prod_{i=1}^N {\cal I}_i\,,
\label{Sgamma}
\\
&&{\cal I}_i=
\th\Big(\Xfun(\min(\Riph,\dzero))-
\sum_{j=1}^i E_j\,\th(\dzero-\Rjph)\Big).
\label{Idef}
\eeqn
It is easy to understand that eq.~(\ref{Sgamma}) is equivalent
to the isolation cuts described above. In particular, the quantity
\mbox{$\prod_{i=1}^N {\cal I}_i$} is equivalent to step 2.
Therefore, ${\cal S}_{\gamma,N}$ is the measurement function relevant
for the isolated-photon plus jets cross section: it vanishes when applied 
to those parton configurations where the photon is non-isolated.

I now turn to the discussion of the main features of the definition 
of isolated photon given in this paper. Clearly, it is 
sufficient here to investigate the behaviour of the cross section
when one or more partons are inside the isolation cone. 
First of all, I observe that for a soft parton ($E\to 0$) 
the isolation cone does not exist at all (see eq.~(\ref{Idef})).
This ensures that the cancellation of
soft gluon effects will take place as in ordinary
infrared-safe jet cross sections. Secondly, eqs.~(\ref{iscond})
and~(\ref{Xfunlim}) imply that a parton is softer the closer 
to the photon axis: a parton exactly collinear to the photon is 
necessarily soft. Thus, when a quark gets collinear to the photon, 
the damping associated with the quark vanishing energy suppresses
the collinear divergence. Therefore, there is no need for
a final-state collinear counterterm.

To be more quantitative, I start by considering the case of a quark
inside the isolation cone. I restrict for the moment to the
case of $\epem$ collisions. The leading behaviour of the 
partonic amplitude squared is \mbox{$1/(1-y)$}, where $y$
is the cosine of the angle between the quark and the photon.
The contribution to the isolated-photon cross section from the
region inside the isolation cone\footnote{Notice that, by integrating
\mbox{$1/(1-y)$} outside the isolation cone, one gets a term proportional
to \mbox{$\log(1-\cos\dzero)$}, as expected in isolated-photon
production.} is therefore
\beq
\sigma_{cone}\sim\int_{\cos\dzero}^{1} dy\int_0 dE\,E\, 
\frac{\th(\Xfun(\delta(y))-E)}{1-y},
\label{sigq}
\eeq
where the factor $E$ originates from the phase space, and the condition 
of eq.~(\ref{iscond}) has been enforced with a theta function. 
The upper integration bound in $E$ is irrelevant in what follows, 
and will be neglected. Using eq.~(\ref{Xfundef}) we get
\beq
\sigma_{cone}\sim\frac{E_\gamma^2\epsilon_\gamma^2}{2}\int_{\cos\dzero}^{1} 
dy\,\frac{1}{1-y}\left(\frac{1-y}{1-\cos\dzero}\right)^{2n}
\,=\,\frac{E_\gamma^2\epsilon_\gamma^2}{4n},
\label{sigmaquark}
\eeq
provided that $n\geq 1/2$. As previously anticipated, the damping 
associated with the energy of the quark which gets soft cancels the 
effects of the collinear divergence, for reasonable choices of $\Xfun$. 
The fact that there is no need for a final-state collinear counterterm
is consistent with the fact that the contribution from the fragmentation 
function is also vanishing; in QCD, fragmentation is a rigorously 
collinear phenomenon, and therefore eq.~(\ref{iscond}) would imply $z=1$.
Thus, the contribution of the fragmentation function is restricted
to a zero-measure set in the phase space.

I now turn to the case of a gluon inside the isolation cone. 
The leading behaviour of the partonic amplitude squared is \mbox{$1/E^2$}. 
Using the subtraction method,\footnote{With the subtraction method, the 
divergent part of the cross section is evaluated strictly in the soft limit, 
$E=0$. As previously stressed, the isolation cone does not constrain 
at all the phase space of soft partons, and therefore this divergent part
will be cancelled by the corresponding virtual contribution as customary 
in perturbative QCD.} the contribution to the finite part of the cross 
section (again, from the region inside the isolation cone) will read
\beq
\sigma_{cone}\sim\int_{\cos\dzero}^{1} dy\int_0 dE\,
\frac{\th(\Xfun(\delta(y))-E)-1}{E}.
\label{sigg}
\eeq
Using eq.~(\ref{Xfundef}) we get
\beq
\sigma_{cone}\sim\int_{\cos\dzero}^{1} dy
\log\left(E_\gamma\epph\left(\frac{1-y}{1-\cos\dzero}\right)^n\right)
\,=\,(1-\cos\dzero)(\log(E_\gamma\epph)-n).
\label{sigmagluon}
\eeq

The case of hadronic collisions is only slightly more complicated.
The $\th$ function in eqs.~(\ref{sigq}) and~(\ref{sigg}) is now
\beq
\th(\Xfun(R(y))-E_{\sss T}).
\eeq
This function clearly constrains the energy of the parton, since
$E=E_{\sss T}\cosh\eta(y)$, where I have explicitly indicated that 
the pseudorapidity of the parton depends upon $y$. In order not to
spoil the conclusions of eqs.~(\ref{sigmaquark}) and~(\ref{sigmagluon}),
one must have \mbox{$E\to 0$} when $y\to 1$, or, which is equivalent,
$\cosh\eta(y)$ must tend to a finite constant when $y\to 1$
(notice that \mbox{$1-\cos R(y)$} tends to zero at the same rate
of \mbox{$1-y$} for $y\to 1$).
This is indeed the case, since for the very definition of $y$ one gets
\mbox{$\cosh\eta(y)\to\cosh\eta_\gamma$}. In isolated-photon production,
the photon is observed in the central region of the detector, and
$\cosh\eta_\gamma$ is of order one.

By increasing the number of partons inside the isolation cone the
situation rapidly becomes rather involved, preventing us to perform
analytical calculations. However, from the definition of the
isolation cuts, it should be clear that, 
given a configuration which fulfills the isolation criteria,
it is always possible that a parton emits a soft gluon, or 
splits into two collinear partons, without changing abruptly
the physical observables. On the other hand, if a configuration
does {\it not} fulfill the isolation criteria, if a parton
emits a soft gluon or splits into two collinear partons, we
get a configuration which still does not fulfill the isolation cuts.
Therefore, any QCD infrared configuration (soft gluons, quarks 
and gluons collinear to each other) can be locally (that is, 
prior to any phase-space integration) subtracted, and the corresponding 
singularities are cancelled by the virtual contributions. The only 
singularities which are left unsubtracted are the QED ones (quarks 
collinear to the photon). However, in this case the singularity is
damped by the mechanism described in eq.~(\ref{sigmaquark}).

The fact that soft gluon emission and collinear splitting do not 
modify the physical observables is formally equivalent to the infrared 
safeness of the cross section to all perturbative orders. However,
it has to be stressed that the isolation cuts have an impact on the
local subtraction of singularities. This can be clearly seen from
eqs.~(\ref{sigg}) and~(\ref{sigmagluon}), where the integration
over the gluon energy $E$, constrained by the isolation cuts, results
in a singular (but integrable) function of $y$ (without any isolation
condition, the integral over $E$ would simply give zero).
It seems reasonable to assume that, at higher orders, the isolation
cuts always define a function which, although singular in some
regions of the phase space, has a finite integral: however, no formal 
proof will be given here. This fact has been explicitly verified in the 
case where a quark and a gluon are inside the isolation cone, which 
is relevant for the next-to-leading order cross section of isolated-photon 
production in $\epem$ collisions.

We can understand the impact of the isolation condition 
on the radiative corrections by looking at eq.~(\ref{sigmagluon}).
The definitions of isolated photon in the cone approach which are
usually adopted in the literature can be recovered by 
setting $n=0$ and $\epph=\epsilon_c$. $\epsilon_c$ 
is the maximum amount of hadronic energy (normalized to the photon energy) 
allowed inside the cone, and {\it to minimize} the contribution of
the fragmentation mechanism it must be a small number.
Therefore, we see from eq.~(\ref{sigmagluon}) that, comparing
to the case where the isolation condition is extremely loose ($n=0$, 
$\epph=1$), there is a sizeable negative correction $\log\epsilon_c$.
On the other hand, with the definition given in this paper, 
$\epph$ does not need to be small. The isolation condition is obtained 
in a smooth way, controlled by the function $\Xfun$. 
From eq.~(\ref{sigmagluon}), the smaller is $n$, the more moderate will 
be the (negative) corrections. This is the reason for the choice 
of $\epph$ and $n$ made in eq.~(\ref{parameters}).
Finally, notice that in standard approaches $\dzero$ is usually 
taken to be of the order of $20^\circ$. With the current 
definition, $\dzero$ can be chosen to be larger. This implies that
the logarithms \mbox{$\log(1-\cos\dzero)$}, which are present 
in the isolated-photon cross section, do not get large.

A final remark is still in order. When $\delta\simeq 0$, 
the value of $\Xfun(\delta)$ is below the energy threshold $E_{th}$ 
of hadronic calorimeters, and therefore the condition of 
eq.~(\ref{iscond}) is not experimentally meaningful any longer.
The data can therefore get a contribution from the fragmentation
process, with \mbox{$z>z_{th}=1-E_{th}/E_\gamma$}. However, from
the experimental point of view there is nothing special in the
region $\delta\simeq 0$ (obviously, there are no singularities
in particle detectors); since $E_{th}$ is typically of the order of 
few hundred MeV, and the photon is required to be hard, $z_{th}\simeq 1$,
and the contribution of the fragmentation mechanism will be 
small. In other words, one could relax eq.~(\ref{iscond})
by allowing quasi-soft partons to be collinear with the photon;
this would imply the presence of a collinear singularity in the
direct cross section, which would be cancelled by a proper 
fragmentation cross section. After the cancellation, one would be left
with a very small finite contribution. By strictly imposing 
eq.~(\ref{iscond}), this small contribution is zero from the beginning,
and the fragmentation cross section is simply not there.

\section{Conclusions}

I presented a definition of isolated-photon plus jets
cross section which is based upon an isolation cone and a 
jet-finding algorithm which excludes the photon. It has been
argued that this prescription, which treats identically quarks
and gluons, is infrared safe to all orders. Hadrons are 
allowed inside the isolation cone, at the border of which they 
can be as energetic as the photon itself, but are required to be 
softer the closer they are to the photon axis, eventually 
becoming soft in the exact collinear limit. This fact 
implies that the contribution of the photons coming from the 
fragmentation of quarks and gluons can be neglected in QCD.
Therefore, for hadron-hadron and photon-hadron collisions, the only 
non-calculable parts which enter the theoretical prediction
are the parton densities of the incoming hadrons. In the case
of $\epem$ collisions, the isolated photon plus jets cross section
is fully calculable in perturbation theory. The prescription given
here is also applicable without any modification to the case
of several, isolated photons in the final state.

\section*{Acknowledgements}
I would like to thank C.~Grab, Z.~Kunszt, M.~Mangano, H.~Niggli, 
S.~Passaggio, G.~Ridolfi and A.~Signer for many discussions
and useful comments. I am especially indebted to S.~Catani, 
who found a mistake in a preliminary version of this paper, 
and to P.~Nason, whose suggestions influenced the final
form of the paper.

\begin{reflist}

\item\label{cone}
 H.~Baer, J.~Ohnemus and J.~F.~Owens, \pr{D42}{90}{61};\\
 P.~Aurenche, R.~Baier and M.~Fontannaz, \pr{D42}{90}{1440};\\
 E.~L.~Berger and J.~Qiu, \pr{D44}{91}{2002};\\
 E.~W.~N.~Glover and W.~J.~Stirling, \pl{B295}{92}{128};\\
 Z.~Kunszt and Z.~Trocsanyi, \np{B394}{93}{139};\\
 L.~E.~Gordon and W.~Vogelsang, \pr{D50}{93}{1901}.
\item\label{demo}
 E.~W.~N.~Glover and A.~G.~Morgan, \zp{C62}{94}{311};\\
 A.~Gerhmann-De~Ridder and E.~W.~N.~Glover, preprint DPT/97/26, 
 \hepph{9707224}.
\item\label{fragfun}
 A.~Gerhmann-De~Ridder, T.~Gerhmann and E.~W.~N.~Glover, 
 \pl{B414}{97}{354}, \hepph{9705305}. 
\item\label{KS}
 Z.~Kunszt and D.~E.~Soper, \pr{D46}{92}{192}.
\item\label{FKS}
 S.~Frixione, Z.~Kunszt and A.~Signer, \np{B467}{96}{399}, \hepph{9512328}.
\item\label{Jets97}
 S.~Frixione, \np{B507}{97}{295}, \hepph{9706545}.

\end{reflist}

\end{document}